\documentclass[10pt,letterpaper,twocolumn]{article} %% two column, final layout
\usepackage{ol}
\usepackage[draft]{hyperref}

\begin{document}

\twocolumn[

\title{Efficient polarization squeezing in optical fibers}

\author{Joel~Heersink, Vincent~Josse, Gerd~Leuchs and Ulrik.~L.~Andersen}
\affiliation{Institut f\"ur Optik, Information und Photonik, Max--Planck Forschungsgruppe, Universit\"at Erlangen--N\"urnberg, G\"{u}nther-Scharowsky-Str. 1, Bau 24, 91058, Erlangen, Germany}

\begin{abstract}

We report on a novel and efficient source of polarization squeezing using a single pass through an optical fiber. Simply passing this Kerr squeezed beam through a carefully aligned $\lambda/2$ waveplate and splitting it on a polarization beam splitter, we find polarization squeezing of up to $5.1 \pm 0.3$~dB. The experimental setup allows for the direct measurement of the squeezing angle.

\end{abstract}

]


\maketitle %% NULL FUNCTION WITH LATEX 2e

The budding field of quantum information holds promise to revolutionize communication. In particular the field of quantum information with continuous variables has received much attention in the last decades with the realization, for instance, of quantum teleportation, quantum cryptography and dense coding.\cite{braunstein.book} Most of these protocols require the use of squeezed (or entangled) states of light that are generally detected using a strong local oscillator, which makes the scalability of quantum networks difficult. Recently, polarization squeezing has attracted much attention \cite{chirkin93.qe,grangier87.prl,bowen02.prl,heersink03.pra,josse03.prl} as it can be measured in direct detection,\cite{korolkova02.pra} making it especially attractive for cryptography and other quantum communication applications. Furthermore the flucuations of the polarization variables can be mapped onto the collective fluctuations of an atomic ensemble, paving the way to quantum memory.\cite{hald99.prl} To date polarization squeezing has been achieved using the nonlinear interactions in Optical Parametric Amplifiers,\cite{bowen02.prl} optical fibers \cite{heersink03.pra} and atomic media.\cite{josse03.prl}

In this paper we present the results of a novel and efficient method of bright pulse polarization squeezing generation using the Kerr effect in an optical fiber. This setup is greatly simplified relative to other schemes and leads to high squeezing levels. Further, this method enables us to completely characterize all quadratures of the state exiting an optical fiber. In particular we measured the angle by which the squeezed uncertainty region has been rotated by the nonlinear Kerr effect.

%\section{Theory}

\begin{figure}[htb]
  \centerline{ \includegraphics[width = 7 cm]{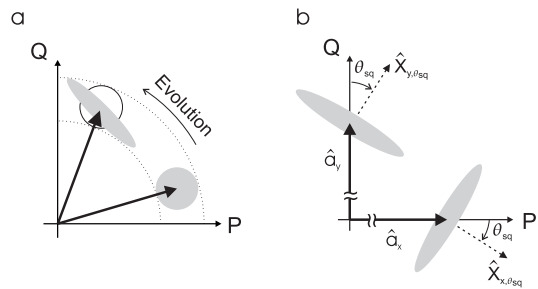}	}
	\caption{Representation in phase space of a) the effect of Kerr squeezing on a coherent beam, and of b) the light state exiting the fiber in our setup.}
	\label{fig:phase_space}
\end{figure}

It has been known for a long time that the optical Kerr effect in glass fiber can generate quadrature squeezing. This third order nonlinear effect ($\chi^3$), also refered to as Self Phase Modulation produces an intensity dependent change in the refractive index. The squeezing generation can be understood by a single mode picture represented in Fig.~\ref{fig:phase_space}(a): since different amplitudes experience different rotations in phase space, the fluctuation circle (corresponding to shot noise) of the input field is transformed into an ellipse. The squeezed quadrature is rotated by the angle $\theta_{sq}$ relative the amplitude quadrature. Since the Kerr effect conserves photon number, the amplitude fluctuations remain at the shot noise level preventing direct detection of the squeezing. %This is a serious limitation because it is not possible to use homodyne detection due to the high intensity of the light beam necessary to obtain a sufficiently high nonlinearity.

%To overcome this problem, one can use an asymmetric Sagnac interferometer \cite{kitagawa86.pra}: mixing an intense quadrature squeezed beam with a weak beam on an higly asymmetric beam splitter, one generates an intense amplitude squeezed beam. This method has been realized with two counterpropagating beams in the same fiber \cite{schmitt98.prl,krylov98.ol}. An elegenat variant of this scheme has been demonstrated \cite{fiorentino01.pra}. Here two beams propagated along the two axes of a polarization maintaining fiber and the asymmetric interference was realized by a half wave plate and polarization beam splitter (PBS). The symmetric Sagnac interferometer, where the squeezing is identical in both arms, provides an alternative \cite{bergman91.ol,rosenbluh91.prl}. By interfering two identical squeezed beams, a squeezed vacuum and an intense quadrature squeezed beam are generated. One major drawback of this scheme is that vacuum squeezing cannot be directly measured.

\begin{figure}[htb]
  \centerline{
    \includegraphics[width = 8 cm]{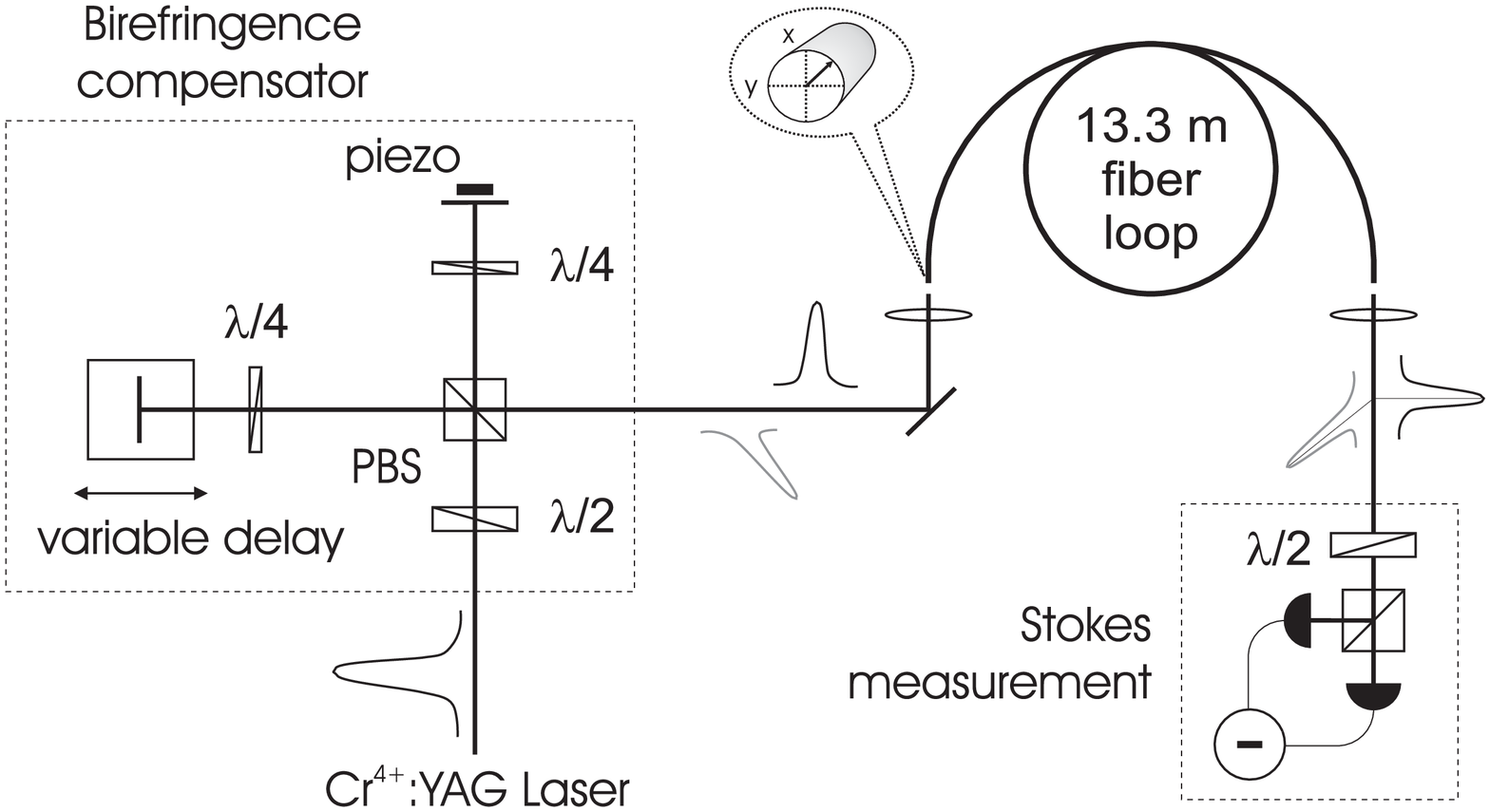}
	}
	\caption{Eexperimental setup for efficient polarization squeezing generation.}
	\label{fig:setup}
\end{figure}

Polarization squeezing overcomes this problem. It can be generated by overlapping two quadrature squeezed beams produced in the two orthogonal polarization modes of the fiber, visualized by $ \hat{a}_{x}$ and $ \hat{a}_{y}$ in Fig.~\ref{fig:phase_space}(b). This setup, seen in Fig.~\ref{fig:setup}, is similar to our previous experiment producing polarization squeezing,\cite{heersink03.pra} where instead two amplitude squeezed beams were generated. These were produced in an asymmetric Sagnac interferometer \cite{kitagawa86.pra,schmitt98.prl,krylov98.ol} in which a strong Kerr squeezed pulse is transformed into amplitude squeezing by interference with a weak auxilliary pulse; squeezing by this principle can also be achieved using a Mach Zehnder interferometer.\cite{fiorentino01.pra} This intereference of a strong squeezed and a weak "coherent" beam however gives rise to a loss in squeezing due to the dissimilarity of the pulses. In this paper's setup we avoid this destructive effect by mutually interfering two strong Kerr squeezed pulses in a Stokes measurement, with the potential to measure more squeezing. Formally this interference of equally squeezed pulses is reminiscent of earlier experiments producing vacuum squeezing.\cite{bergman91.ol,rosenbluh91.prl}  Further advantages of the present setup is a greater robustness against input power fluctuations and the ability to measure squeezing at all powers. 

%We present here an improved version of the Sagnac interferometer where the two identical quadrature squeezed beams, $ \hat{a}_{x}$ and $ \hat{a}_{y}$ in Fig.~\ref{fig:phase_space}(b), are generated in a single pass through an optical fiber, much as in \cite{fiorentino01.pra}. We however mutually interfere the two beams emerging from the fiber's two polarization axes (Fig.~\ref{fig:setup}) thereby conserving all correlations. In analogy to the symmetric Sagnac fiber loop \cite{bergman91.ol,rosenbluh91.prl} where two identical squeezed beams are interfered to produce vacuum squeezing, a squeezed vacuum could be separated at our fiber output with, for example, a set of waves plates and a polarization beam splitter (PBS). However, retaining this two mode superposition leads to polarization squeezing. Compared with the asymmetric Sagnac interferometer \cite{kitagawa86.pra,schmitt98.prl,krylov98.ol}, which has also been used to generate polarization squeezing \cite{heersink03.pra}, we have a simplified setup which exhibits greater correlations visible at all powers.

\begin{figure}[htb]
  \centerline{
    \includegraphics[width = 8.2 cm]{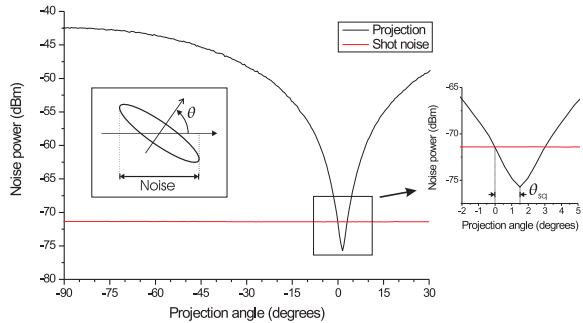}
	}
	\caption{Noise against phase-space rotation angle for the rotation of the measurement $\lambda /2$ waveplate for a pulse energy of 83.7~pJ using 13.3~m 3M~FS-PM-7811 fiber. Inset: Schematic of the projection principle for angle $\theta$. Results are corrected for $-86.1\pm0.1$~dBm electronic noise.}
	\label{fig:results_rotation}
\end{figure}

To describe the quantum polarization state of a light field it is helpful to introduce the quantum Stokes parameters.\cite{korolkova02.pra} These are defined in analogy to their classical counterparts:
\begin{eqnarray}
	\hat{S}_{0} &= \hat{a}^{\dagger}_{x} \hat{a}_{x} + \hat{a}^{\dagger}_{y} \hat{a}_{y}, \quad
	\hat{S}_{1} &= \hat{a}^{\dagger}_{x} \hat{a}_{x} - \hat{a}^{\dagger}_{y} \hat{a}_{y},  \nonumber \\
	\hat{S}_{2} &= \hat{a}^{\dagger}_{x} \hat{a}_{y} + \hat{a}^{\dagger}_{y} \hat{a}_{x}, \quad
	\hat{S}_{3} &= i(\hat{a}^{\dagger}_{y} \hat{a}_{x} - \hat{a}^{\dagger}_{x} \hat{a}_{y}).
	\label{eqn:stokes_def}
\end{eqnarray}
Following the non-commutation of the photon annihilation and creation operators, $\hat{a}_{x/y}$ and $\hat{a}^{\dagger}_{x/y}$, these Stokes parameters obey the relations: $ [\hat{S}_0,\hat{S}_i] = 0$ and $ [\hat{S}_i,\hat{S}_j] = 2i\epsilon_{ijk}\hat{S}_k$ with $ \{i,j,k \}= 1,2,3$. These relations lead to Heisenberg inequalities for the fluctuations of these parameters that depend on the mean polarization state.\cite{korolkova02.pra} For instance, let us consider the situation where the modes $ \hat{a}_{x}$ and $\hat{a}_{y}$ have the same amplitude but phase shifted by $\pi/2$ (Fig.~\ref{fig:phase_space}(b)): $\left\langle \hat{a}_{x} \right\rangle=i\left\langle \hat{a}_{y} \right\rangle=i\alpha/\sqrt{2}$, $\alpha$ being a real number. This light is circularly polarized ($\langle\hat{S}_1\rangle=\langle\hat{S}_2\rangle=0$; $\langle\hat{S}_3\rangle=\langle\hat{S}_0\rangle=\alpha^2$) and the only non-trivial Heisenberg inequality is $\Delta^{2} \hat{S}_{1}\, \Delta^{2} \hat{S}_{2}\geq{}|\langle\hat{S}_3\rangle|^{2}= \alpha^{4}$, where $\Delta^{2} \hat{S}_j$ refers to the variance $\langle\hat{S}_j^2\rangle - \langle\hat{S}_j\rangle^2$. Polarization squeezing is achieved if the variance $\Delta^2\hat{S}_{\theta}$ (variance of a general Stokes parameter rotated by an angle $\theta$ in the $\hat{S}_1$-$\hat{S}_2$ plane) is below the shot noise level: 
\begin{equation}
	\Delta^{2} \hat{S}_{\theta} \! \leq{} \! |\langle \hat{S}_{3} \rangle | \! = \! \alpha^{2} \quad{} \textnormal{where} \quad{} \hat{S}_{\theta} \! = \! \cos{\theta} \, \hat{S}_1 \! + \! \sin{\theta} \, \hat{S}_2.
	\label{eqn:stokes_uncertainty}
\end{equation}
To fully characterize the polarization fluctuations, one can express the fluctuations of the Stokes parameters in terms of the noise of the linearly polarized modes $ \hat{a}_{x}$ and $\hat{a}_{y}$. Since the fluctuations are small compared to the mean values ($\delta \hat{a}_x,\,\delta \hat{a}_y\,<<\, \alpha$), we find:
 \begin{eqnarray}
	 %\delta\hat{S}_{1}&= &\alpha (\delta\hat{X}_{x}^{+}-\delta\hat{X}_{y}^{+})/\sqrt{2}\\
	 %\delta\hat{S}_{2}&=& \alpha (\delta\hat{X}_{x}^{-}-\delta\hat{X}_{y}^{-})/\sqrt{2}\\
	 \delta\hat{S}_{\theta}&= &\alpha (\delta\hat{X}_{x,\theta}-\delta\hat{X}_{y,\theta})/\sqrt{2}.
	\label{eqn:stokes_linearization}
\end{eqnarray}
Here $\hat{X}_{x/y, \theta}$ corresponds to the quadrature rotated by an angle $\theta$ for $x$ and $y$ polarizations, visualized in Fig.~\ref{fig:phase_space}(b) and Fig.~\ref{fig:results_rotation}. The amplitude and phase quadratures are found for $\theta = 0$ and $\theta = \pi/2$.

%, which are defined as $\hat{X}_{x/y, \theta}= i(e^{i\theta}\hat{a}^{\dagger}_{x/y}-e^{-i\theta}\hat{a}_{x/y})$. 
%correspond to $\hat{X}_{x/y}^{+}=\hat{X}_{x/y, 0}$ and $\hat{X}_{x/y}^{-}=\hat{X}_{x/y, \pi/2}$. 

The modes $x$ and $y$ propagate through the same fiber with identical amplitudes. They experience the same nonlinearity, and thus have the same quadrature noise $\Delta^{2} \hat{X}_{x,\theta}=\Delta^{2} \hat{X}_{y,\theta}\equiv\Delta^{2} \hat{X}_{\theta}$. We assume the noise to be uncorrelated since the modes are temporally and spatially separated by the fiber birefringence. With this assumption, the measured noise corresponds to the noise of the rotated quadrature, $\Delta^2 S_{\theta} = \alpha^2 \Delta^2 X_{\theta} $. Polarization squeezing is seen in a certain quadrature, rotated by the angle $\theta_{sq}$. This leads to two conclusions: i) we can easily produce polarization squeezing using this setup and, ii) by measuring the fluctuations of the Stokes parameters we can characterize the light state, in particular its quadrature noise $\Delta^2 X_{\theta}$.    
 
%\section{Experiment}

The experimental setup is seen in Fig.~\ref{fig:setup}. We use a Cr$^{4+}$:YAG laser emitting 130~fs FWHM pulses at 1497~nm at a repetition rate of 163~MHz. This beam is coupled into the two orthogonal polarization axes of a 13.3~m polarization maintaining fiber (3M FS-PM-7811, 6~$\mu$m core diameter). In this manner we generate two independent Kerr squeezed beams in a single pass through a fiber. After the fiber, the pulses' intensities are adjusted to be identical and they are aligned to temporally overlap. The fiber's polarization axes exhibit a strong birefringence (beat length 1.67~mm) that must be compensated. To minimize losses at the fiber output, we precompensate the pulses in an unbalanced Michelson-like interferometer that introduces a tunable delay between the polarizations.\cite{heersink03.pra,fiorentino01.pra} At the output, 0.1\% of the light is used as an input signal for an control loop which maintains a constant relative phase of $\pi/2$ between the exiting pulses, producing a circularly polarized beam (not explicitly shown).

We measure the noise of a given Stokes parameter with the half waveplate $(\lambda /2)$ and a PBS. The PBS outputs are measured directly using detectors with Epitaxx-500 photodiodes and a low pass filter ($\leq$40~MHz) to avoid AC saturation due to the laser repetition rate. The measurement frequency is 17.5~MHz on a HP8595E spectrum analyzer with a resolution bandwidth of 300~kHz and a video bandwidth of 30~Hz. The difference between the two photocurrents is given by:
\begin{equation}
	 i_- \; \propto \; \cos{4\Phi}\,\hat{S}_1\!+\!\sin{4\Phi}\,\hat{S}_2\;=\;\hat{S}_{4\Phi},
	\label{eqn:stokes_measurements}
\end{equation}
where $\Phi$ is the rotation angle of the waveplate compared to the $x$ axis. Rotating the waveplate effectively rotates the measured Stokes parameter in the $\hat{S}_1$-$\hat{S}_2$ plane, allowing a complete measurement of the polarization noise. Using the fact that the amplitude fluctuations of the two individual modes $x$ and $y$ are at the shot noise level, the Heisenberg uncertainty limit is determined by measuring $\Delta^2\hat{S}_1$. This shot noise calibration was checked using a coherent beam with the same amplitude directly from the laser.  
%At high power and longer fiber lengths we find that the squeezing is optimized for unequal powers in the two polarization axes. Thus, the MUS for our $\hat{S}_1$-$\hat{S}_2$ plane is calculated by measuring the optical power in two orthogonal (i.e. squeezing and anti-squeezing) bases in this plane, and subtracting this from the total optical power.

%\section{Results}

A plot of the measured noise as the $\lambda /2$ waveplate is rotated is seen in Fig.~\ref{fig:results_rotation}. The pulse energy is 83.7~pJ (soliton energy 56$\pm$4~pJ). We find an oscillation between very large noise and squeezing, as expected from the rotation of a squeezed state. Plotted on the x-axis is the projection angle $\theta$, i.e. the angle by which the state has been rotated in phase space, inferred from the waveplate angle ($\theta=4\Phi$). For $\theta=0$, an $\hat{S}_1$ measurement, we find a noise value equal to the shot noise. Rotation of the state by $\theta_{sq}$ makes the squeezing in the system observable by projecting out only the squeezed axis of the uncertainty ellipse. Further rotation brings a rapid increase in noise as the excess phase noise, composed of the anti-squeezing and classical thermal noise, becomes visible. This is similar to experiments using local oscillators, however here no stabilization is needed after production of the polarization squeezed state. This may be important for experiments with long acquisition times, i.e. state tomography. %To gain in precision this angle was actually measured by determining the difference between the two waveplate angles for which a level equal the shot noise was observed and dividing this value in half.

\begin{figure}[htb]
  \centerline{
    \includegraphics[width = 7 cm]{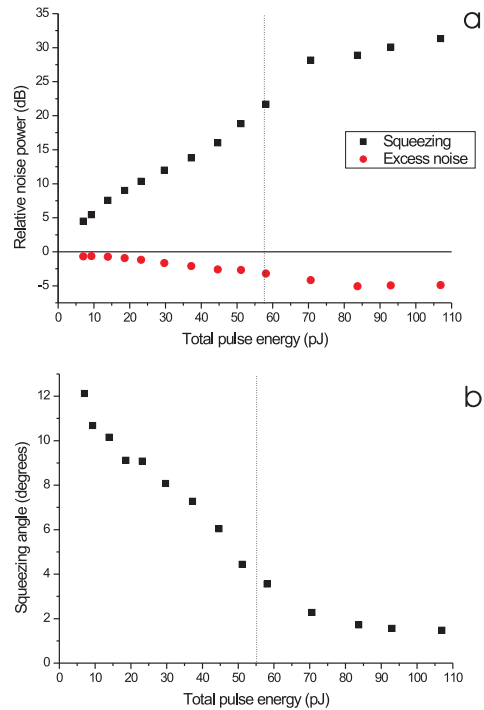}
	}
	\caption{Results for 13.3~m 3M FS-PM-7811 fiber as a function of pulse energy: a) the squeezing and excess phase noise and b) the squeezing angle. The energy at which a first order soliton is generated (56$\pm$4~pJ) is shown by the dashed line. Results are corrected for $-86.1\pm0.1$~dBm electronic noise.}
	\label{fig:results_13.3m}
\end{figure}

The squeezed and anti-squeezed quadratures and the squeezing angle, $\theta_{sq}$, of this state were investigated as a function of pulse energy (Fig.~\ref{fig:results_13.3m}). The maximum observed squeezing is $-5.1\pm0.3$~dB at an energy of 83.7~pJ.
%The squeezing was checked by optically attenuating the beam and measuring the squeezing; the resultant curve was linear indicating true squeezing. 
Squeezing saturation is seen at high power, likely due to the overwhelming excess phase noise which distorts the uncertainty ellipse. The losses of the setup were found to be 20.5\%: 4\% from the fiber end, 7.8\% from optical elements and 10\% from the photodiodes. Thus we have generated a maximum polarization squeezing of $-8.8\pm0.8$~dB. This value agrees better with theoretical predictions.\cite{drummond01.josab} Investigating the squeezing angle, $\theta$, we find that the rotation of the uncertainty region necessary to observe squeezing decreases with increasing power. This is expected as, despite an increasing anti-squeezing, the amplitude noise of a Kerr squeezed beam remains constant. Saturation is also apparent in $\theta_{sq}$ making a clean projection of the squeezing difficult. This polarization squeezing setup could be further investigated for different fiber lengths as well as types and theoretical description of the setup could be implemented.

%\section{Conclusion}

We see potential for further improvement of our $-5.1\pm0.3$~dB squeezing produced in our novel and efficient setup, namely with better photodiodes and an all fiber setup in which losses are minimized. The developement of specialty microstructured fibers with lower classical phase noise is expected to also improve our result,\cite{korn04.iqec} bringing us yet closer to the theoretically predicted maximal fiber squeezing.\cite{drummond01.josab} Polarization and quadrature entanglement can be directly produced from this resource, both important tools in many quantum communication protocols.

%We have investigated a novel and efficient source of polarization squeezing using a single pass through a glass fiber. With this setup we were able to directly measure the squeezing angle as well as excellent polarization squeezing of $-5.1\pm0.3$~dB.

%\section*{Acknowledgments}

This work was funded by Project 1078 of the DFG. The authors thank O. Gl\"{o}ckl for help. U.L.A. thanks the Alexander von Humboldt Foundation.

%
% Reference List
%

\end{document}